\documentclass[12pt]{article}
\usepackage{geometry}
\usepackage{caption}
\usepackage{subcaption}
\usepackage{braket}
\usepackage{amsmath, amsfonts}
\usepackage{mathptmx} 	
\usepackage{graphicx, array}
\usepackage{hyperref}
\usepackage[all]{hypcap}
\usepackage{float}
\usepackage{longtable}
\usepackage[utf8]{inputenc}
\usepackage{algorithm,algpseudocode}
\hypersetup{
    colorlinks=true,
    citecolor=blue,
    linkcolor=blue,
    filecolor=magenta,      
    urlcolor=blue,
}
\usepackage{qcircuit}
\usepackage{authblk}

\begin{document}

\title{Preparation of quantum superposition using partial negation}
           
\author[1,2]{Sara Anwer\thanks{Sara.Anwar@alexu.edu.eg}}
\author[1,2]{Ahmed Younes\thanks{ayounes@alexu.edu.eg}}
\author[1,2]{Islam Elkabani\thanks{islam.kabani@alexu.edu.eg}}
\author[1,2]{Ashraf Elsayed\thanks{ashraf.elsayed@alexu.edu.eg}}
\affil[1]{Department of Mathematics and Computer Science, Faculty of Science, Alexandria University, Alexandria,Egypt}
\affil[2]{Alexandria Quantum Computing Group, Faculty of Science, Alexandria University, Alexandria, Egypt.}

\date {}
\maketitle
\abstract
The preparation of a quantum superposition is the key to the success of many quantum algorithms and quantum machine learning techniques. The preparation of an incomplete or a non-uniform quantum superposition with certain properties is a non-trivial task. In this paper, an $n$-qubits variational quantum circuit using partial negation and controlled partial negation operators will be proposed to prepare an arbitrary quantum superposition. The proposed quantum circuit follows the symmetries of the unitary Lie group. The speed of the preparation process and the accuracy of the prepared superposition has a special importance to the success of any quantum algorithm. The proposed method can be used to prepare the required quantum superposition in $\mathcal{O}(n)$ steps and with high accuracy when compared with relevant methods in literature.
\section*{keyword}
Quantum superposition; Quantum state; partial negation; data encoding; prepared amplitudes; acquired amplitudes

\section{Introduction}
Quantum computing \cite{1} is a technology that uses the properties of quantum mechanics, including entanglement \cite{2} and superposition \cite{3,7}.
The quantum computers can solve optimization problems faster than classical computers, by using quantum laws of entanglements and superposition\cite{3,7}.
Qubit is the basic data unit in quantum computer where each qubit can hold states $\ket 0$ and $\ket 1$ simultaneously \cite{2}, and this is called quantum superposition. 
The preparation of quantum superposition is a prerequisite stage in building many quantum algorithms in several domains \cite{4}. For example, in data encoding \cite{28}, quantum machine learning \cite{29}, the artificial neural networks (ANN)\cite{5}, Grover’s Search algorithm \cite{8,9}, quantum Fourier transform \cite{17}, quantum linear system algorithms \cite{23}, quantum Image processing \cite{10}, etc.

The problem of preparing a superposition of quantum states over n-qubits using given amplitudes has been tackled in many previous work \cite{24,26,27,12,19,25,16}.
In \cite{24}, a quantum circuit was described for the preparation of quantum state distributions, however, it used exponential number of elementary gates. The work in \cite{26} developed the quantum circuit in \cite{24} by using quantum multiplexers in order to reduce the total number of elementary gates. This technique was important for the preparation of incomplete superpositions, however, the number of gates were still of exponential complexity. A quantum circuit was introduced in \cite{27} to reduce the number of CNOT control gates in \cite{24} from $2^n$ to $\frac{23}{24}2^n$ by using quantum universal gates.
In \cite{16}, a quantum circuit was designed in order to prepare \emph{Prime States}. These states are highly entangled for n-qubits and used in the twin primes distribution applications, where it only prepares \emph{Prime} distributions.
A quantum circuit was proposed in \cite{12} to  prepare an equal superposition state by applying the Hadamard gate or the $Y$ operator on each qubit, where $Y$ is a rotation with $\theta = \pi/2$ angle along the $y$ axis, where it only prepares \emph{Equal States} distributions.
In \cite{25}, two different approaches of quantum state preparation were introduced. In the first approach, a sequential algorithm was proposed to construct a quantum circuit with of exponential number of gates having linear number of auxiliary qubits. The second approach suggested a parallel algorithm to build a quantum circuit with polynomial number of gates and exponential number of auxiliary qubits. 
A circuit optimization technique was presented in \cite{19} in order to reduce the complexity of the state preparation circuit by using basic numerical integration. The total number of gates in this circuit is linear with the number of qubits, however, it also uses a linear number of auxiliary qubits.

The aim of this paper is to propose an n-qubits variational quantum circuit based on partial negation operators for preparing an arbitrary quantum superposition, where these operators follows the symmetries of the unitary Lie group \cite{22}. Given the required quantum superposition (vector of amplitudes) to be prepared, the parameters of the variational circuit are calculated numerically using Levenberg-Marquardt algorithm by transforming the vector of amplitudes to a system of non-linear equations. The number of gates in the proposed quantum circuit is linear with the number of qubits and is able to prepare both complete and incomplete superpositions with high accuracy.

The remainder of this paper is organized as follows. Section \ref{2} provides an overview on the standard formulation of the partial negation operator that will be used in the proposed quantum circuit.  Section \ref{3} explains the proposed method. Section \ref{4} presents and discusses the experimental of the proposed method. Section \ref{5} compares the complexity of this method with other related work in literature. Finally, Section \ref{c6} concludes the paper.

\section{The Partial Negation Operator}\label{2}
In order to prepare a superposition using the proposed variational quantum circuit, it is required to provide a background about the partial negation operator used by this method.   

The $X$ gate \cite{3} is the quantum gate which is equivalent to the classical NOT gate and is represented as follows:
\begin{equation}
	X =
	\begin{bmatrix}
		0 & 1\\
		1 & 0 
	\end{bmatrix}.
\end{equation}
The partial negation operator $K$ is the $r^{th}$ root of the $X$ gate \cite{13} and can be calculated using the following equation,
\begin{equation}
	K =\sqrt[r]{X}=\frac{1}{2}
	\begin{bmatrix}
		1+s & 1-s\\
		1-s & 1+s 
	\end{bmatrix},
\end{equation}
where $s=\sqrt[r]{-1}$ is a parameter that decides the behavior of the gate (the degree of negation).

The $K$ gate will be used to introduce the operator $C_{k}$. The $C_{k}$  is an operator on $n+1$ qubits register that applies $K$ conditionally for $n$ times on an auxiliary qubit denoted as $\ket{ak}$ and initialized to state $\ket0$.  The number of times the $K$ gate is applied on $\ket{ak}$ is based on the number of qubits. In general, $C_{k}$ can be represented as follows,
\begin{equation}
	C_k = Cont\_K(x_0; ak)Cont\_K(x_1; ak) . . . Cont\_K(x_{n-1}; ak),
\end{equation}
where the $Cont\_K(x_i; ak)$ gate is a 2-qubits controlled gate with control qubit $\ket{x_i}$ and target qubit $\ket{ak}$. The $Cont\_K(x_{i}; ak)$ gate applies $K$ conditionally on $\ket{ak}$ if $\ket{x_i} =\ket 1$, so when $C_k$ is applied on vector $\ket{x_0 x_1 . . . x_{(n-1)}}$ and  $\ket{ak} = \ket0$ can be understood as follows,
\begin{equation}
	C_k(\ket{x_0 x_1 . . . x_{(n-1)}}\otimes \ket0)=\ket{x_0 x_1 . . . x_{(n-1)}}\otimes (\dfrac{1+s}{2}\ket0+\dfrac{1-s}{2}\ket1).
\end{equation}

Finally, in order to measure the probabilities of finding the auxiliary qubit $\ket {ak}$ in state $\ket {0}$ or $\ket {1}$, the following equations are used

\begin{equation}
	\begin{split}
		Pr(\ket{ak}=\ket0)=\mid{\dfrac{1+s}{2}}\mid^2, \\
		Pr(\ket{ak}=\ket1)=\mid{\dfrac{1-s}{2}}\mid^2.
	\end{split}
\end{equation}

\section{Preparation of Quantum Superposition}\label{3}
In this section, the general quantum state method used in the preparation of the quantum superposition is presented. The general form of any quantum superposition can be presented by the following equation:
\begin{equation}\label{6}
	\ket{\psi}	= \sum_{i=0}^{2^n-1} a_i \ket i,
\end{equation}
where $n$ is the number of qubits and $a_i$ is the amplitudes of each state. A state $\ket i$ can be represented as a binary form $\ket {i_n .....i_1}$, where $ i \in  \{0,1\}^n$ for $n$ qubits.\\ 
The main idea of the method is to find the parameters for the quantum circuit that prepares a superposition by introducing a novel approach for calculating the parameters of its gates. The $r^{th}$ root gates are adopted in the design of the circuit which contains $n$ qubits in addition to an auxiliary qubit  $\ket{ak}$ in order to store the basis states. This method represents the circuit as a system of non-linear equations whose variables are the unknown parameters of the gates. Finally, a method for evaluating the accuracy of the method is described. However, we will first describe in the following subsection the process of construct the quantum circuit used in the proposed method. 
\\
\subsection{Construction of the Variational Quantum Circuit }  \label{prep.cir}
\begin{figure}
	\[
	\Qcircuit @C=0.5em @R=0.5em @!R{
		\lstick{\ket{x_0}} & \qw & \gate{K_1} &\qw & \ctrl{4} & \qw & \qw &\qw & \qw & \qw & \qw & \qw\\
		\lstick{\ket{x_1}}  & \qw& \gate{K_2} &\qw &   \qw &  \qw &\ctrl{3} & \qw  &\qw & \qw & \qw & \qw\\
		\vdots    & & & & & & & \cdots\\
		\lstick{\ket{x_{n-1}}} & \qw & \gate{K_n} &\qw &  \qw &  \qw &\qw &\qw & \qw &\ctrl{1} & \qw & \qw\\
		\lstick{\ket{0}} & \qw &  \qw  &\qw &  \gate{K_{n+1}} &  \qw  &\gate{K_{n+2}} &\qw & \qw &  \gate{K_{2n}} & \qw & \qw& &{\ket{ak}}
	}
	\]
	\caption{General quantum circuit for the proposed method.}
	\label{fig1}
\end{figure}
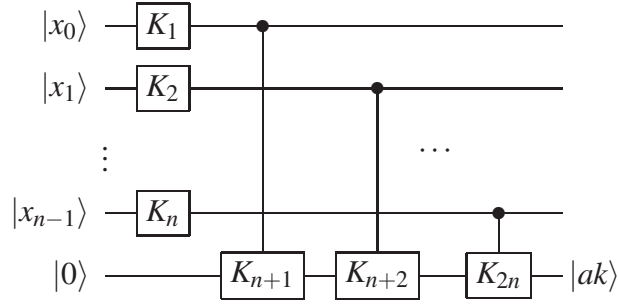

In order to construct the quantum circuit, a quantum register of $n+1$ qubits all initialized to the state $\ket{0}$ in the form

\begin{equation}
	\ket{\psi_0}	=	\ket0^{\otimes {n+1}}.
\end{equation}
By applying the $K_t$ gates on the first $n$ qubits, the following superposition is produced
\begin{equation}
	W_1	= K_1\otimes K_2\otimes K_3 \otimes \dots \otimes K_n \otimes I,
\end{equation} 
where $K_t$ could be any partial negation operator with any $r^{th}$ root and $t$ is an index of the gate in range $\{1, \dots ,2n \}$. Now, when applying $W_1$ on $\ket{\psi_0}$, the following state will be produced
\begin{eqnarray}
	\ket{\psi_1} &=& W_1\ket{\psi_0}\nonumber \\
	&=&   K_1\otimes K_2\otimes K_3 \otimes \dots \otimes K_n \otimes I \ket{0}.\\ \nonumber
\end{eqnarray}
Finally, the $n$ two qubits controlled operator $C_k$ is applied on the targets of the auxiliary qubit $\ket{ak} $ which have a control on each qubit for the first $n$ qubits. This will produce the following state
\begin{eqnarray}\label{eq13}
	\ket{\psi_2} &=& C_k\ket{\psi_1} \\ \nonumber
	&=&(\prod_{l=1}^{n} Cont\_k(x_l;ak)\otimes I^{\otimes {n-1}} )( K_1\otimes K_2\otimes K_3 \otimes \dots \otimes K_n \otimes I) \ket{0}^{\otimes {n+1}}\\\nonumber
	&=& a_1 \ket{000..0} + a_2 \ket{000...1} +.....+a_{2^n}\ket{011...1}, \nonumber
\end{eqnarray}
where $a_i$ is the value of the amplitudes, as shown in fig. \ref{fig1}.
\\
\makeatletter
\newenvironment{breakablealgorithm}
{
	\begin{center}
		\refstepcounter{algorithm}
		\hrule height.8pt depth0pt \kern2pt
		\renewcommand{\caption}[2][\relax]{
			{\raggedright\textbf{\ALG@name~\thealgorithm} ##2\par}%
			\ifx\relax##1\relax 
			\addcontentsline{loa}{algorithm}{\protect\numberline{\thealgorithm}##2}%
			\else 
			\addcontentsline{loa}{algorithm}{\protect\numberline{\thealgorithm}##1}%
			\fi
			\kern2pt\hrule\kern2pt
		}
	}{
		\kern2pt\hrule\relax
	\end{center}
}
\makeatother
\subsection{The Proposed Method}
In this section the proposed method for preparing the superposition of any quantum system of maximum 3-qubits will be described in details. The input for this method is the values of the exact amplitudes of some distribution represented as a vector of size $2^n$ of the form $(a_1,a_2, \dots ,a_{2^n})$, where $n$ is the number of qubits and $a_{i} $ for $i\in {1, 2, \dots, 2^n}$ is an arbitrary complex number. The output of the method are the parameters of $r^{th}$ root gates in the form 
\begin{equation}\label{mat}
	K_t=\sqrt[r]{X}=\frac{1}{2}
	\begin{bmatrix}
		1+s & 1-s\\
		1-s & 1+s 
	\end{bmatrix}
	=
	\begin{bmatrix}
		c_{2m-1} & c_{2m}\\
		c_{2m} & c_{2m-1} 
	\end{bmatrix},
\end{equation}
where $m \in \{1,\dots,4n\},c_{2m-1} = \frac{1}{2} (1+s)$ and $c_{2m} =\frac{1}{2} (1-s)$ are the unknown parameters of gates for  $s=\sqrt[r]{-1}$. The method will achieve this by solving a system of nonlinear equations for the unknown parameters of the gates. This system of equations is generated by eq. ($\ref{eq13}$) for any $n$ qubits with maximum $3$ qubits.  
The system of non-linear equations can be represented as 
\begin{eqnarray} \label{eq14}
	a_1&=&\prod_{l=0}^{n} c_{2(n-l)-1},\label{eqa1}\\
	a_2&=&(\prod_{l=0}^{n} c_{2n-3l+\alpha})(c_{4n-1}+c_{4n})\label{eqa2} ,\\
	a_3&=&(\prod_{l=0}^{n} c_{n-l^2+2})(c_{3n}+c_{3n+1})\label{eqa3} ,\\
	a_4&=&(\prod_{l=0}^{n} c_{2(n-l)-\alpha} )( \sum_{l=3n}^{3n+1} \sum_{j=4n-1}^{4n}c_lc_j),\label{eqa4} \\
	a_5&=&(\prod_{l=0}^{n-1} c_{n+l^2-\alpha-1})(c_{2n+1}+c_{2(n+1)}),\label{eqa5} \\
	a_6&=&(\prod_{l=0}^{n-1} c_{n+l^2-1})(\sum_{l=2n+1}^{2(n+1)}\sum_{j=4n-1}^{4n}c_lc_j),\\
	a_7&=&(\prod_{l=0}^{n} c_{2n-l^2-\alpha-1})(\sum_{l=2n+1}^{2(n+1)}\sum_{j=3n}^{3n+1}c_lc_j) \label{eqa7} ,\\
	a_8&=&(\prod_{l=0}^{n} c_{2(n-1)})(\sum_{l=2n+1}^{2(n+1)}\sum_{j=3n}^{3n+1}\sum_{j=4n-1}^{4n}c_lc_j) \label{eqa8},
\end{eqnarray}
where $a_i$ are the values of the exact amplitudes and $c_i$ are the unknown parameters of the gate $K_t$, such that $c_y = 1 $ for $y \leq 0 $.  For the eqs. (\ref{eqa2}),(\ref{eqa4}),(\ref{eqa5}) and (\ref{eqa7}), the variable $\alpha=1$ for $l=2$ and $\alpha=0$ otherwise.
This system of equations represents the case of 3-qubits. For a single qubit case eqs.(\ref{eqa1}) and (\ref{eqa2}) will be used while for 2-qubits case eqs. (\ref{eqa1}) to (\ref{eqa4}) will be used. This system of non-linear equations is solved using Levenberg-Marquardt algorithm \cite{15}.

The complete method proposed in this paper is presented in algorithm ~\ref{alg1}. The algorithm starts by preparing an $n+1$ qubit quantum circuit as described in section \ref{prep.cir}. A list, namely \emph{AcquiredGates}, is used to store the parameters of the gates resulting from solving the system of non-linear equations eqs. (\ref{eqa1}) to (\ref{eqa8}). These gates are then applied on $\ket{\phi}$ to produce a new vector of amplitudes stored in the \emph{AquiredAmplitude} list. Finally, in order to measure the accuracy of the proposed method, the relative error between the acquired probability and the prepared probability is calculated using eq. (\ref{eq16}). The probability of the prepared amplitudes and the probability of  the acquired amplitudes are given by eqs. (\ref{exact}) and (\ref{acquired}), respectively.
\begin{equation} \label{exact}
	Pr\_prepared=\mid a \mid^2,
\end{equation}
where $a$ is the prepared amplitudes.
\begin{equation}\label{acquired}
	Pr\_acquired= \mid a' \mid^2,
\end{equation}
where $a'$ is the acquired amplitudes.\\

The relative error is the difference between the prepared probability and the acquired probability is given by the following  equation.

\begin{equation}\label{eq16}
	relativeError=\dfrac{\mid(Pr\_prepared-Pr\_acquired)\mid}{Pr\_prepared}.
\end{equation}
\begin{breakablealgorithm}
	\caption{Quantum State Preparation Algorithm}
	\label{alg1}
	\begin{algorithmic}[1]	
		\noindent{Given a system $\ket{\psi}$,	and the vector of amplitudes $(a_1,a_2, \dots ,a_{2^n})$}
		\State  Initialize $\ket{ak}$ in state $\ket{0}$ 
		\State  Prepare $\Ket{\phi} = (\ket{\psi}\otimes\ket{ak})$
		\State Let \emph{AcquiredAmplitude} = $a'$= []
		\State Let \emph{AcquiredGates}=[]
		\State Apply $K$ with unknown parameters on $\ket{\psi}$
		\State Then apply $C_k$ on $\ket{\phi}$ 
		\State The amplitudes of $\ket{\phi}$ is required to solve the system of equations  
		\State Solve the system of non-linear equations generated by steps 1 to 7 using Levenberg-Marquardt algorithm
		\State Save the result from the system in \emph{AcquiredGates}
		\State Apply  the gates on the state $\Ket{\phi}$ 
		\State Calculate the \emph{AcquiredAmplitude} 
		\State Measure the accuracy of the method by using the relative error 	
	\end{algorithmic}
\end{breakablealgorithm}

The parameters of the gates resulting from algorithm \ref{alg1} must represent unitary gates $(UU^\dagger=I)$ in order to preserve the reversibility of the resulting quantum gates. To verify the  reversibility of the gates the following unitary test must be applied:
\begin{equation}\label{eq17}
	\begin{split}
		\sum_{i=1}^{4n} \mid c_{2i-1}\mid^2+\mid c_{2i}\mid^2 -1 = 0\\
	\end{split}
\end{equation}
\begin{equation}
	\begin{split}
		\sum_{i=1}^{4n} c_{2i-1}c_{2i}^*+c_{2i}c_{2i-1}^* = 0 ,
	\end{split}
\end{equation}
where $c_i^*$ is the complex conjugate transpose of $c_i$.\\
The following section shows a detailed example to illustrate the steps of algorithm \ref{alg1} and its can be applied for the case of 3-qubits as well as showing the resulting the unitary test is satisfied for the resulting gates.
\\
\noindent

\subsection{Detailed Example} \label{alg}
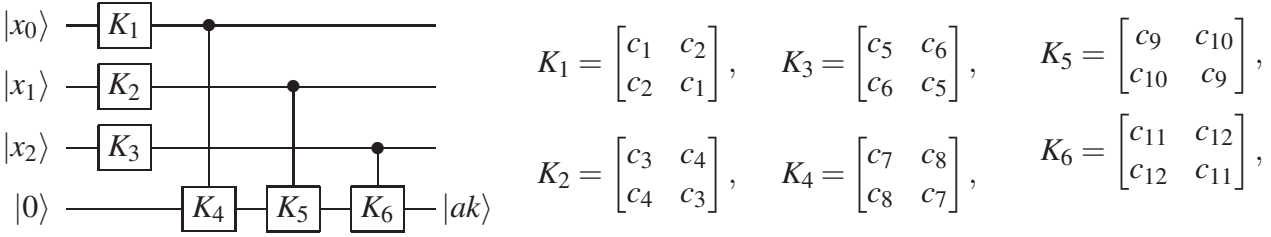
\begin{figure}[h]
	\begin{tabular}{>{\centering\arraybackslash}m{6cm}m{3cm}m{2cm}m{5cm}} 
		
		\[
		\Qcircuit @C=0.5em @R=0.5em @!R{
			\lstick{\ket{x_0}} & \qw & \gate{K_1} &\qw & \ctrl{3} & \qw & \qw &\qw & \qw & \qw & \qw\\
			\lstick{\ket{x_1}}  & \qw& \gate{K_2} &\qw &   \qw &  \qw &\ctrl{2} & \qw  &\qw & \qw & \qw\\
			\lstick{\ket{x_2}} & \qw & \gate{K_3} &\qw &  \qw &  \qw &\qw &\qw & \ctrl{1} & \qw & \qw\\
			\lstick{\ket{0}} & \qw &  \qw  &\qw &  \gate{K_4} &  \qw &\gate{K_5} &\qw &  \gate{K_6} & \qw & \qw& &{\ket{ak}}
		}
		\]
		&
		\begin{equation}
			K_1=
			\begin{bmatrix}
				c_1 & c_2\nonumber \\
				c_2 & c_1 		
			\end{bmatrix},
		\end{equation}
		\begin{equation}
			K_2=
			\begin{bmatrix}
				c_3 & c_4\nonumber \\
				c_4 & c_3 		
			\end{bmatrix},
		\end{equation}
		& \begin{equation}
			K_3=
			\begin{bmatrix}
				c_5 & c_6\nonumber \\
				c_6 & c_5 
			\end{bmatrix},
		\end{equation}
		\begin{equation}
			K_4=
			\begin{bmatrix}
				c_7 & c_8\nonumber \\
				c_8 & c_7 
			\end{bmatrix},
		\end{equation}
		&\begin{equation}
			K_5=
			\begin{bmatrix}
				c_9 & c_{10}\nonumber \\
				c_{10} & c_9 
			\end{bmatrix},
		\end{equation}
		\begin{equation}
			K_6=
			\begin{bmatrix}
				c_{11} & c_{12}\nonumber \\
				c_{12} & c_{11} 
			\end{bmatrix},
		\end{equation}
	\end{tabular}
	\caption{quantum circuit with unknown $r^{th}$ roots for $3$ qubits.}\label{figg2} 	
\end{figure}

In this example a quantum circuit is prepared with a quantum register of $3$ qubits and one auxiliary qubit  $\ket{ak}$, all initialized with state $\ket0$, as follows,
\begin{equation}
	\ket{\psi_0}	=	\ket0^{\otimes4},
\end{equation}
then the operators $K_1,K_2 $ and $ K_3$ are applied on each qubit as shown in fig. \ref{figg2}, to produce the state $\ket{\psi_1}$ as follows
\begin{eqnarray}
	\ket{\psi_1} &=& K_1\otimes K_2\otimes K_3 \otimes I \ket{\psi_0}.
\end{eqnarray}
where $ r \in \{1,2,3\}$. 
Finally, the state $\ket{\psi_1}$ is produced by applying operators $C_k$ taking the first $3$ qubits as control $\ket{x_i}$ and $\ket{ak}$ is the target qubit for each operator $C_k$.The following is the resulting state:

\begin{eqnarray} \label{eq30}
	\ket{\psi_2} &=&C_k\ket{\psi_1} \\ \nonumber
	&=& Cont\_K(x_0; ak)Cont\_K(x_1; ak) Cont\_K(x_2; aK)(K_1\otimes K_2\otimes K_3 \otimes I )\ket0^{\otimes4}\\ \nonumber
	&=&c_1c_3c_5\ket{0000}+c_1c_3c_6(c_{11}+c_{12})\ket{0001}+\dots+c_2c_4c_6(c_7c_9c_{11}+ c_7c_{10}c_{11}\\ \nonumber &&+c_7c_9c_{12} +c_7c_{10}c_{12}+c_8c_9c_{11}+c_8c_{10}c_{11}+c_8c_9c_{12}+c_8c_{10}c_{12})\ket{0111}. \\ \nonumber
\end{eqnarray}

The $c_i$ values, for $i\in \{1,2,\dots,12\}$, resulting from eq.( \ref*{eq30})  are the values of the amplitudes representing the entries of the gates as shown in fig. \ref{figg2}.

For example, assume an arbitrary vector of amplitudes: $a=[-0.1500+0.5100i,0.4400 + 0.1200i, 0.3680 + 0.1110i, 0.0900 - 0.3200i,0.2920 + 0.0920i,0.0760 - 0.2500i, 0.0610 - 0.2130i, \\-0.1830 - 0.0510i] $given as input, the system of non-linear equations in eq. (\ref{eq18}) is produced. The values of the \emph{AcquiredGates} are calculated by solving this system of non-linear equations.

\begin{alignat}{2}\label{eq18}
a_1 &=-0.1500+0.5100i & &=c_1c_3c_5,\nonumber\\
a_2 &=0.4400 + 0.1200i & &=c_1c_3c_6(c_{11}+c_{12}), \nonumber\\
a_3 &= 0.3680 + 0.1110i & &=c_1c_4c_5(c_9+c_{10}),\nonumber\\
a_4 &= 0.0900 - 0.3200i & &=c_1c_4c_6 (c_9c_{11}+c_9c_{12}+c_{10}c_{11}+c_{10}c_{12}),\nonumber\\
a_5 &=0.2920 + 0.0920i & &=c_2c_3c_5(c_7+c_8), \nonumber\\
a_6 &=0.0760 - 0.2500i & &=c_2c_3c_6(c_7c_{11}+c_7c_{12}+c_8c_{11}+c_8c_{12}) ,\nonumber\\
a_7 &=0.0610 - 0.2130i & &=c_2c_4c_5(c_7c_9+c_7c_{10}+c_8c_9+c_8c_{10}) ,\nonumber\\
a_8 &=-0.1830 - 0.0510i & &=c_2c_4c_6(c_7(c_9c_{11}+c_{10}c_{11}+c_9c_{12}+c_{10}c_{12})+c_8(c_9c_{11}+
c_{10}c_{11}+c_9c_{12}+c_{10}c_{12}) ),\nonumber\\
\end{alignat}

where $c_i$ is the complex number of the parameters of the gate $K_t$ .  
Applying this system on $3$ qubits returns $6$ gates.  To verify the correctness of the parameters of the gates, the unitary test must be applied. eq. (\ref{eqn17}) is an example of applying the unitary test on the $K_1$ gate and its complex conjugate transpose $K_1'$.
\begin{equation}\label{eqn17}
	K_1=\begin{bmatrix}
		0.7784 + 0.3818i& 0.2192 - 0.4468i\\0.2192 - 0.4468i & 0.7784 + 0.3818i \end{bmatrix}
	K_1'=\begin{bmatrix}0.7784 - 0.3818i &0.2192 + 0.4468i\\0.2192 + 0.4468i & 0.7784 - 0.3818i\end{bmatrix}
\end{equation}
where $c_1=0.7784 + 0.3818i, c_2 =	0.2192 - 0.4468i$, the parameters of the gate $K_1$, are the results from solving the system of eq. (\ref{eq18}). Obviously, the value from $K_1K_1^\dagger = I$, where $I$ is the unitary gate. This implies that the unitary test is satisfied. Now, when applying the resulting gates after passing the unitary test on the circuit in fig. \ref{figg2}, the results are $8$ basis states from $\ket{0000}$ to $\ket{0111}$ with acquired amplitudes values [$-0.1503 + 0.5103i,0.4404 + 0.1222i,0.3698 + 0.1089i,0.0885 - 0.3190i,0.2918 + 0.0900i,0.0733 - 0.2519i,0.0652 - 0.2114i,-0.1825 - 0.05\\31i$]. Finally, to measure the accuracy of the method, the eqs. (\ref{exact}) and (\ref{acquired}) are calculated using the prepared and acquired amplitudes to produce the following prepared and acquired probabilities  [$0.2826,0.2080,0.1477,0.1105,0.0937,0.0683,0.0491,0.0361$] and [$ 0.2831,0.2088,0.1486,\\0.1096,0.0933,0.0688,0.0490,0.0361$], respectively. The relative error calculated between these 2 vectors using eq. (\ref{eq16}) was 
$7.5342\times10^{-4}$. 

The following section discusses other test cases with different values of amplitudes.

\section{Experimental Results and Discussion}\label{4}
This section presents the results from applying some test cases adopted for evaluating the proposed method and also comparing it with related algorithms in literature  \cite{12}, \cite{16} and \cite{19}. First we discuss the results for test cases used in preparing superpositions for $n=3$ qubits and one auxiliary qubit, where all qubits are initialized with state $\ket 0$. The prepared amplitudes for $7$ different test cases are used in the experiments. After each experiment, the relative error between the prepared amplitudes and the acquired amplitudes are calculated. Table \ref{table} summarizes the results for the $7$ test cases.  
\\
As shown in table \ref{table}, the different distribution states are represented as prepared amplitudes.
The first 2 cases were tested against related algorithms  \cite{12} and \cite{16} with given real and complex amplitudes. The other test cases were suggested  to cover the large spectrum of different distributions. More than $100$ random test cases, in addition to the $7$ adopted test cases, were used to evaluate the accuracy of the proposed method. In general, the range of the relative error for all test cases was between $6.9593\times 10^{-11}$ and $0.0987$. The maximum relative error occurred when the prepared amplitudes were represented with real numbers, however, when it was represented using complex numbers, the relative error was less than 0.0001 in various test cases such as the\emph{ Decreasing }and\emph{ Increasing States}.  The minimum relative error occurred with the case of \emph{Equal States} prepared amplitudes. This implies that a large spectrum of distributions can be prepared accurately using the proposed method. 

\begin{center}
\begin{longtable}{p{2.6cm}|p{3.5cm}|p{3.55cm}|p{2.35cm} } 

 	\hline  	\label{table}
	
	States & Prepared Amplitudes  &Acquired Amplitudes  & Relative Error\\
	\hline
	\emph{Equal States} (complex values) \cite{12} & $-0.2500+0.2500i\ket{0}$ $\,+0.2500+0.2500i\ket{1}$ $\,+0.2500+0.2500i\ket{2}$ $\,+0.2500-0.2500i\ket{3}$ $\,+0.2500+0.2500i\ket{4}$ $\,+0.2500-0.2500i\ket{5}$ $\,+0.2500-0.2500i\ket{6}$ $\,-0.2500$ $\,$ $\,-$ $\,$ $\, 0.2500i\ket{7}$ &$-0.2500 + 0.2500i \ket{0}$ $\,+0.2500 + 0.2500i\ket{1}$ $\,+0.2500 + 0.2500i\ket{2}$ $\,+0.2500 - 0.2500i\ket{3}$ $\,+0.2500 + 0.2500i\ket{4}$ $\,+0.2500 - 0.2500i\ket{5}$ $\,+0.2500 - 0.2500i\ket{6}$ $\,-0.2500 $ $\,$ $\,-$ $\,$ $\, 0.2500i\ket{7}$ &$ 6.9593\times10^{-11}$\\ [1ex] 
	\hline
	\emph{Equal States}(real values) & $\frac{1}{\sqrt{8}}(\ket{0}+\ket{1}+\ket{2}$ $\,+\ket{3}+\ket{4}+\ket{5}+\ket{6}$ $\,+\ket{7})$ &$-0.0202 + 0.0087i \ket{0}$ $\,-0.0487 + 0.0949i\ket{1}$ $\,-0.0487 + 0.0949i\ket{2}$ $\,+0.1111 + 0.5053i\ket{3}$ $\,+0.0060 + 0.0196i\ket{4}$ $\,+0.0830 + 0.0550i\ket{5}$ $\,+0.0830 + 0.0550i\ket{6}$ $\,+0.4804$ $\,$ $\, -$ $\,$ $\, 0.0510i\ket{7}$ &$ 0.0093$\\ [1ex] 
	\hline
	\emph{Prime States} (complex values) \cite{16} &$0.50i\ket{2}-0.50i\ket{3}$ $\,+0.50i\ket{5}$ $\,$ $\,+$ $\,$ $\,$ $\,0.50i\ket{7}$ &  $ 0.0231 + 0.0444i\ket{0}$ $\,-0.1495 + 0.1137i\ket{1}$ $\,+ 	0.1518 - 0.0419i\ket{2}$ $\,+0.2524 + 0.5345i \ket{3}$ $\,+0.0156 - 0.0488i\ket{4}$ $\,+0.1903 + 0.0265i\ket{5}$ $\,-0.1399-0.0799i \ket{6}$ $\,+0.2059$ $\,$ $\,-$ $\,$ $\,$ $\,0.5688i\ket{7}$
	&0.0054 \\ [1ex]  
	\hline  
	\emph{Prime States} (real values) &$\frac{1}{\sqrt{4}}(\ket{2}+\ket{3}+\ket{5}+\ket{7})$ &  $ -0.089\ket{0}+0.207i\ket{1}$ $\,+ 0.259i\ket{2}+0.600 \ket{3}$ $\,+ 0.075i\ket{4}+	0.175\ket{5}$ $\,+0.219 \ket{6}$ $\,$ $\,+$ $\,$ $\,0.507i\ket{7}$
	&0.0461 \\ [1ex]  
	\hline 
	\emph{Decreasing States} (complex values) &$-0.1500 + 0.5100i\ket{0}$ $\,+0.4400 + 0.1200i\ket{1}$ $\,+ 0.3680 + 0.1110i\ket{2}$ $\,+0.0900 - 0.3200i\ket{3}$ $\,+0.2920 + 0.0920i\ket{4}$ $\,+ 0.0760 - 0.2500i\ket{5}$ $\,+0.0610 - 0.2130i\ket{6}$ $\,-0.1830$ $\,$ $\, -$ $\,$ $\, 0.0510i\ket{7}$ 
	
	&$-0.1503 + 0.5103i\ket{0}$ $\,+0.4404 + 0.1222i\ket{1}$ $\,+ 0.3698 + 0.1089i\ket{2}$ $\,+0.0885 - 0.3190i\ket{3}$ $\,+0.2918 + 0.0900i\ket{4}$ $\,+ 0.0733 - 0.2519i\ket{5}$ $\,+0.0652 - 0.2114i\ket{6}$ $\,-0.1825$ $\,$ $\, -$ $\,$ $\, 0.0531i\ket{7}$ & $7.5342\times10^{-4}$ \\ [1ex]
	\hline
	\emph{Decreasing States} (real values) &$\frac{1}{\sqrt{2}}\ket{0}+\frac{1}{\sqrt{4}}\ket{1}$ $\,+\frac{1}{\sqrt{8}}\ket{2}+\frac{1}{\sqrt{16}}\ket{3}$ $\,+\frac{1}{\sqrt{32}}\ket{4}+\frac{1}{\sqrt{64}}\ket{5}$ $\,+\frac{1}{\sqrt{128}}\ket{6}$ $\,$ $\,$ $\,+$ $\,$ $\, \frac{1}{\sqrt{256}}\ket{7}$  &$-0.255 + 0.6968i\ket{0}$ $\,+
	0.1488 + 0.5803i\ket{1}$ $\,+
	0.0113 + 0.0855i\ket{2}$ $\,+
	0.0466 + 0.0518i\ket{3}$ $\,
	-0.0017 + 0.0385i\ket{4}$ $\,+
	0.0164 + 0.0264i\ket{5}$ $\,+
	0.0019 + 0.0041i\ket{6}$ $\,+
	0.00310$ $\, + $ $\,$ $\,0.0018i\ket{7}$
	&  0.0402 \\ [1ex] 
	\hline
	\emph{Increasing States} (complex values) &$-0.1830 - 0.0510i\ket{0}$ $\,+0.0610 - 0.2130i\ket{1}$ $\,+0.0760 - 0.2500i\ket{2}$ $\,+0.2920 + 0.0920i\ket{3}$ $\,+0.0900 - 0.3200i\ket{4}$ $\,+ 0.3680 + 0.1110i\ket{5}$ $\,+0.4400 + 0.1200i\ket{6}$ $\,-0.1500$ $\,$ $\, +$ $\,$ $\, 0.5100i\ket{7}$   
	&$-0.182- 0.0507i\ket{0}$ $\,+0.0631 - 0.2108i\ket{1}$ $\,+0.0713 - 0.2515i\ket{2}$ $\,+0.2914 + 0.0887i\ket{3}$ $\,+0.0869 - 0.3185i\ket{4}$ $\,+ 0.3691 + 0.1083i\ket{5}$ $\,+0.4402 + 0.1223i\ket{6}$ $\,-0.1524$ $\,$ $\, + $ $\,$ $\,0.5101i\ket{7}$ & $2.5487\times10^{-4}$ \\ [1ex]
	\hline
	\emph{Increasing States} (real values) & $\frac{1}{\sqrt{256}}\ket{0}+\frac{1}{\sqrt{128}}\ket{1}$ $\,+\frac{1}{\sqrt{64}}\ket{2}+\frac{1}{\sqrt{32}}\ket{3}$ $\,+\frac{1}{\sqrt{16}}\ket{4}+\frac{1}{\sqrt{8}}\ket{5}$ $\,+\frac{1}{\sqrt{4}}\ket{6}$ $\,$ $\,$ $\,$ $\,$ $\,+$ $\,$ $\,$ $\,$ $\,$ $\,\frac{1}{\sqrt{2}}\ket{7}$  &$-0.0026 + 0.0035i\ket{0}$ $\,+
	-0.0008 + 0.0056i\ket{1}$ $\,+
	0.0011 + 0.0338i\ket{2}$ $\,+
	0.0219 + 0.0385i\ket{3}$ $\,
	-0.0607 + 0.0385i\ket{4}$ $\,
	-0.0469 + 0.0817i\ket{5}$ $\,
	-0.1869 + 0.5309i\ket{6}$ $\,+
	0.1071$ $\,$ $\, + $ $\,$ $\,$ $\,0.7294i\ket{7}$ & 0.0225 \\ [1ex] 
	\hline
	\emph{Even States} (complex values) &$0.50i(\ket{0}+\ket{2}+\ket{4}+\ket{6})$  &$ 0.498i\ket{0}+0.499\ket{2}$ $\,+ 0.499\ket{4}$ $\,$ $\,-$ $\,$ $\, 0.502i\ket{6}$ & $4.3354\times10^{-6}$ \\ [1ex] 
	\hline
	\emph{Even States} (real values)&$\frac{1}{\sqrt{4}}(\ket{0}+\ket{2}+\ket{4}+\ket{6})$  &$ 0.0137\ket{0}+0.1001\ket{2}$ $\,+ 0.1001\ket{4}+$ $\,0.7322\ket{6}$ & $0.0363$ \\ [1ex] 
	\hline
	\emph{Odd States} (complex values) & $0.50i(\ket{1}+\ket{3}+\ket{5}+\ket{7})$ &$ 0.499i\ket{1}+0.499\ket{3}$ $\,+0.499\ket{5}$ $\,$ $\,-$ $\,$ $\, 0.501i\ket{7}$ & $1.0413\times10^{-6}$ \\ [1ex] 
	\hline
	\emph{Odd States} (real values) & $\frac{1}{\sqrt{4}}(\ket{1}+\ket{3}+\ket{5}+\ket{7})$ &$ 0.0122\ket{1}+	0.0937\ket{3}$ $\,+0.0937\ket{5}+$ $\,0.7204\ket{7}$ & $ 0.0191$ \\ [1ex] 
	\hline
	\emph{Random States} (complex values) &$0.0220 + 0.6000i\ket{0}$ $\,+0.3440 - 0.0130i\ket{1}$ $\,+0.6000 - 0.0200i\ket{2}$ $\,-0.0200 - 0.3450i\ket{3}$ $\,+ 0.1320 - 0.0050i\ket{4}$ $\,-0.0030 - 0.0790i\ket{5}$ $\,-0.0060 - 0.1370i\ket{6}$ $\,-0.0790 $ $\,$ $\,+ $ $\,$ $\,0.0030i\ket{7}$ 
	
	&$0.0205 + 0.5994i\ket{0}$ $\,+0.3446 - 0.0159i\ket{1}$ $\,+0.6004 - 0.0219i\ket{2}$ $\,-0.0167 - 0.3451i\ket{3}$ $\,+0.1351 - 0.0048i\ket{4}$ $\,-0.0037 - 0.0777i\ket{5}$ $\,-0.0052 - 0.1353i\ket{6}$ $\,-0.0778 $ $\,$ $\,+$ $\,$ $\, 0.0039i\ket{7}$  &$ 6.4361\times10^{-5}$ \\ [1ex]
	\hline
	\emph{Random States} (real values) &$0.6004\ket{0}$ $\,+0.3442\ket{1}$ $\,+0.6003\ket{2}$ $\,+0.3456\ket{3}$ $\,+0.1321\ket{4}$ $\,+ 0.0791\ket{5}$ $\,+0.1371\ket{6}$ $\,+0.0791\ket{7}$ 	
	&$-0.2544 + 0.6107i\ket{0}$ $\,-0.0015 + 0.0833i\ket{1}$ $\,+	0.1056 + 0.6711i\ket{2}$ $\,+0.0435 + 0.0737i\ket{3}$ $\,-0.0061 + 0.0282i\ket{4}$ $\,+	0.0006 + 0.0036i\ket{5}$ $\,+0.0098 + 0.0279i\ket{6}$ $\,+0.0024$ $\,$ $\, +$ $\,$ $\,$ $\, 0.0028i\ket{7}$  &$ 0.0987$ \\ [1ex]
	\hline
\caption{The relative error in superposition preparation for $3$-qubit circuits in different distributions.}
\end{longtable}
\end{center}

The \emph{Equal State} test case has been adopted for different quantum algorithms, such as Fourier transform \cite{17}. It is represented using the following equation \begin{equation}
	\ket{\psi_{equal}} = \frac{1}{\sqrt{2^n}}\sum_{i}^{2^n-1}\ket{i},
\end{equation}where $i$ is the basis state, and $n$ is the number of qubits. The resulting parameters of the gates after applying the algorithm has produced acquired amplitudes that are approximately equal to the prepared complex amplitudes  (relative error = $6.9593\times 10^{-11}$), but the relative error =$0.0093$ when prepares \emph{Equal State} with real amplitudes.

The \emph{Prime State} distribution is well known of being highly entangled and helped in the encoding of many theoretical functions such as the distribution of twin primes \cite{16}. It is represented as \begin{equation}
	\ket{\psi_{prime}} = \frac{1}{\sqrt{\pi(2^n)}}\sum_{i\in prime<2^n}\ket{i},
\end{equation}  where $\pi(2^n)$ is the amplitude of the number of \emph{prime States} between $[0,2^n]$. This test case gives a $0.0098$ relative error when prepares the complex amplitudes and the relative error =$0.0461$ for prepared real amplitudes.

Two other special test cases have been introduced in order to evaluate the proposed method, namely,\emph{ the Decreasing and Increasing }distributions. Both test cases used in many applications, such as in risk analysis \cite{20}. The distribution of \emph{the Decreasing} test case is represented by 
\begin{equation}
	\ket{\psi_{dec}}=\sum_{i=0}^{2^n-1}\frac{1}{\sqrt{2^{i+1}}}\ket{i},
\end{equation} 
whereas the \emph{Increasing} test case is represented by 
\begin{equation}
	\ket{\psi_{inc}}=\sum_{i=0}^{2^n-1}\frac{1}{\sqrt{2^{2^{n}-i}}}\ket{i},
\end{equation}

The relative error for both test cases are less than $10^{-4}$ by using prepared complex amplitudes. However, the relative error is less than $10^{-2}$ for preparing real amplitudes.

Another two special distributions adopted as test cases in this paper are the \emph{Even} and\emph{ Odd States} distributions. These two distributions have many applications in the domain of numerical integration \cite{21}.
The \emph{Even} test case is represented by the equation  
\begin{equation}
	\ket{\psi_{even}}	= \frac{1}{\sqrt{2^n/2}}\sum_{i\in \{0,2,\dots,2^n-2\}} \ket {i},
\end{equation}
and the \emph{Odd} distribution test case is represented by 
\begin{equation}
	\ket{\psi_{odd}}	= \frac{1}{\sqrt{2^n/2}}\sum_{i\in \{1,3,\dots,2^n-1\}} \ket {i}.
\end{equation}  The resulting relative error for these two test cases are less than $10^{-6}$ when the values of the prepared amplitudes are complex values and less than $10^{-2}$ when the values of the prepared amplitudes are real amplitudes.

All other test cases adopted for evaluating the method are randomly generated based on eq.(\ref{6}). The calculated average relative error for these test cases was $6.9912\times 10^{-04}$. The result of one of these test cases is presented in table \ref{table}. All the Random test cases gives the relative error $6.4361\times10^{-5}$ by using complex values of amplitudes and the relative error is $0.0987$ for real values of amplitudes.

\begin{figure}
	\centering
	\begin{subfigure}[b]{0.3\textwidth}
		\centering
		\includegraphics[height=2in]{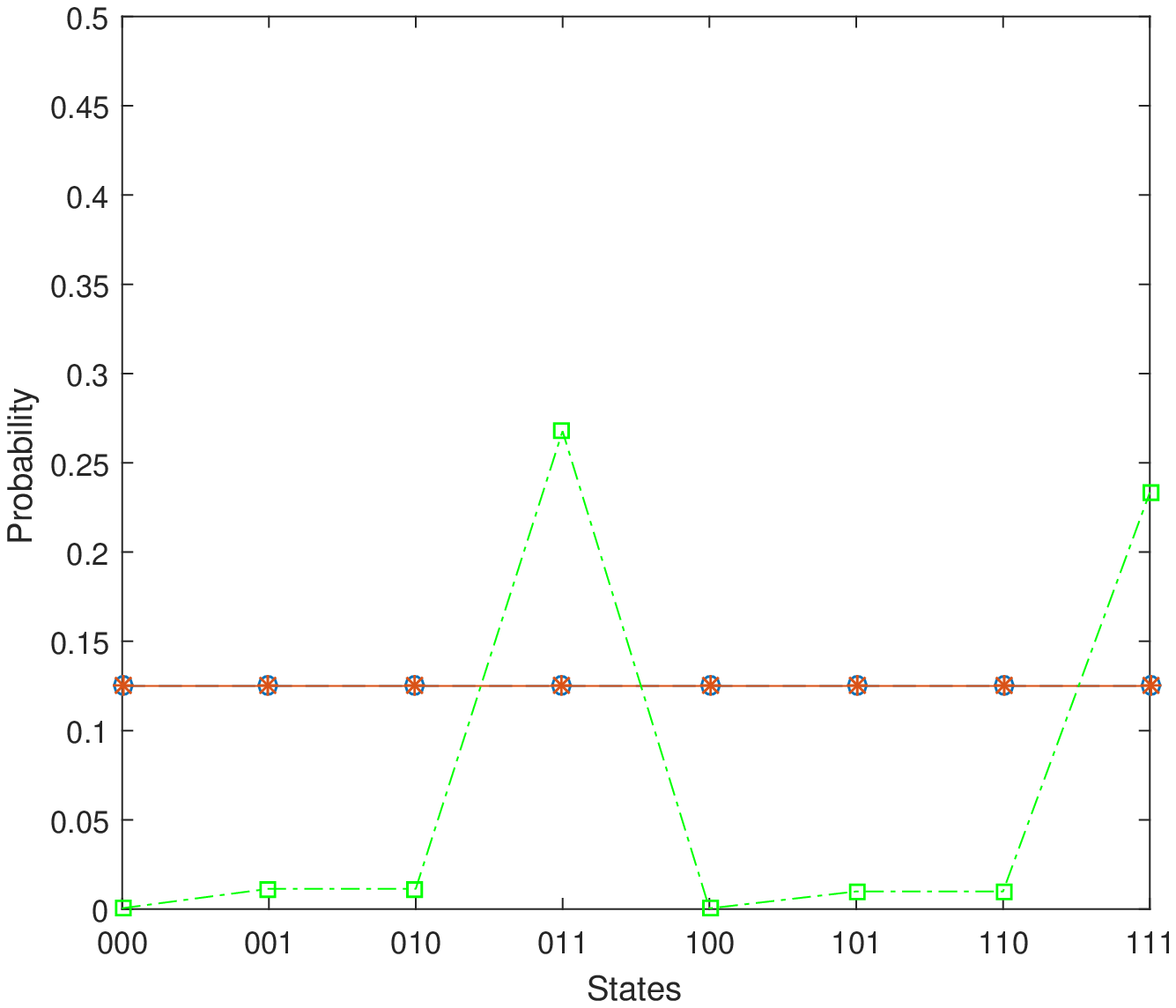}
		\caption{\emph{Equal State}\\}
		\label{fig:Ra}
	\end{subfigure}
	\hfill
	\begin{subfigure}[b]{0.3\textwidth}
		\centering
		\includegraphics[height=2in]{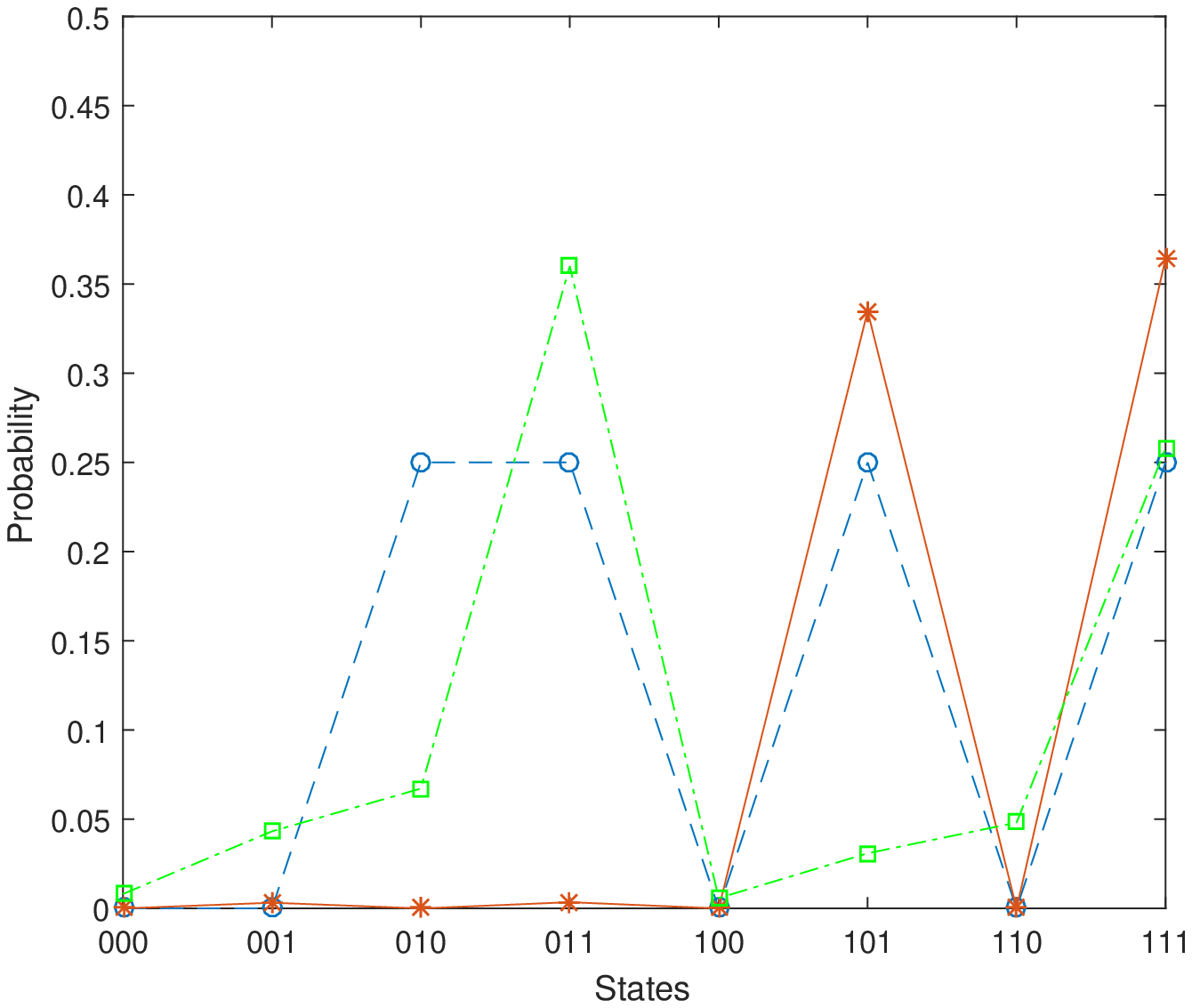}
		\caption{\emph{Prime State}\\}
		\label{fig:Rb}
	\end{subfigure}
	\hfill
	\begin{subfigure}[b]{0.3\textwidth}
		\centering
		\includegraphics[height=2in]{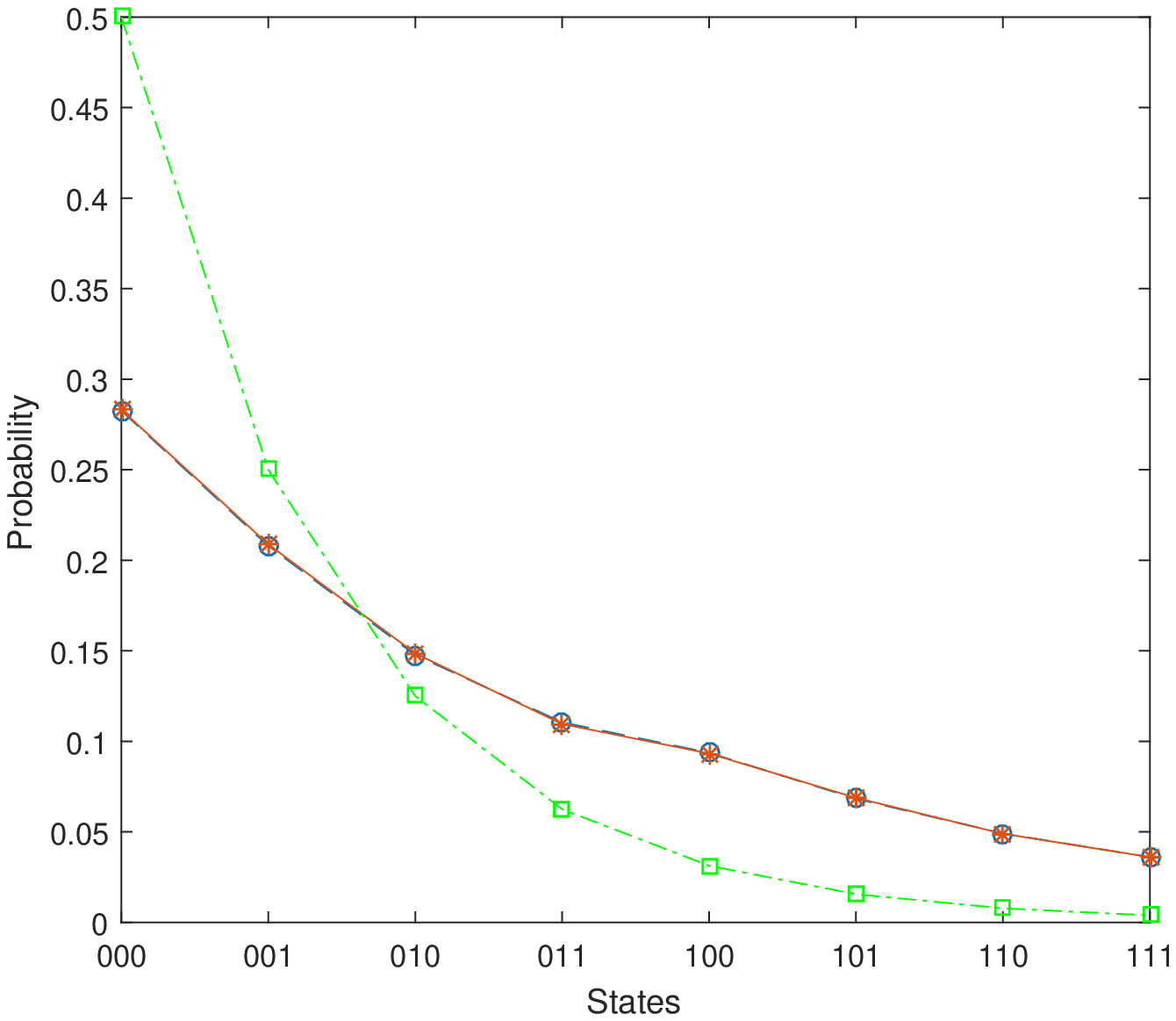}
		\caption{\emph{Decreasing State}\\}
		\label{fig:Rc}
	\end{subfigure}
	\hfill
	\begin{subfigure}[b]{0.3\textwidth}
		\centering
		\includegraphics[height=2in]{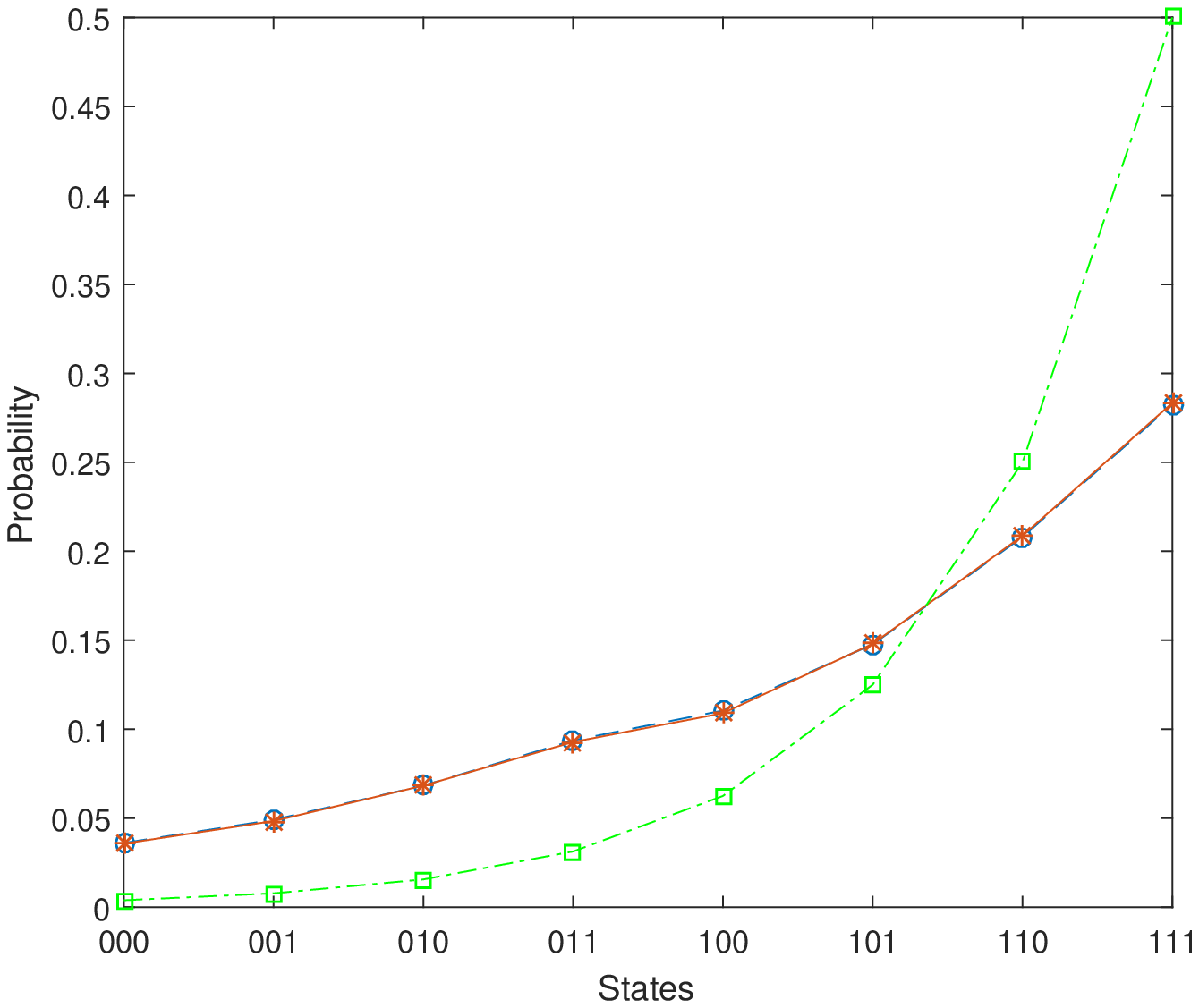}
		\caption{\emph{Increasing State}\\}
		\label{fig:Rd}
	\end{subfigure}
	\hfill
	\begin{subfigure}[b]{0.3\textwidth}
		\centering
		\includegraphics[height=2in]{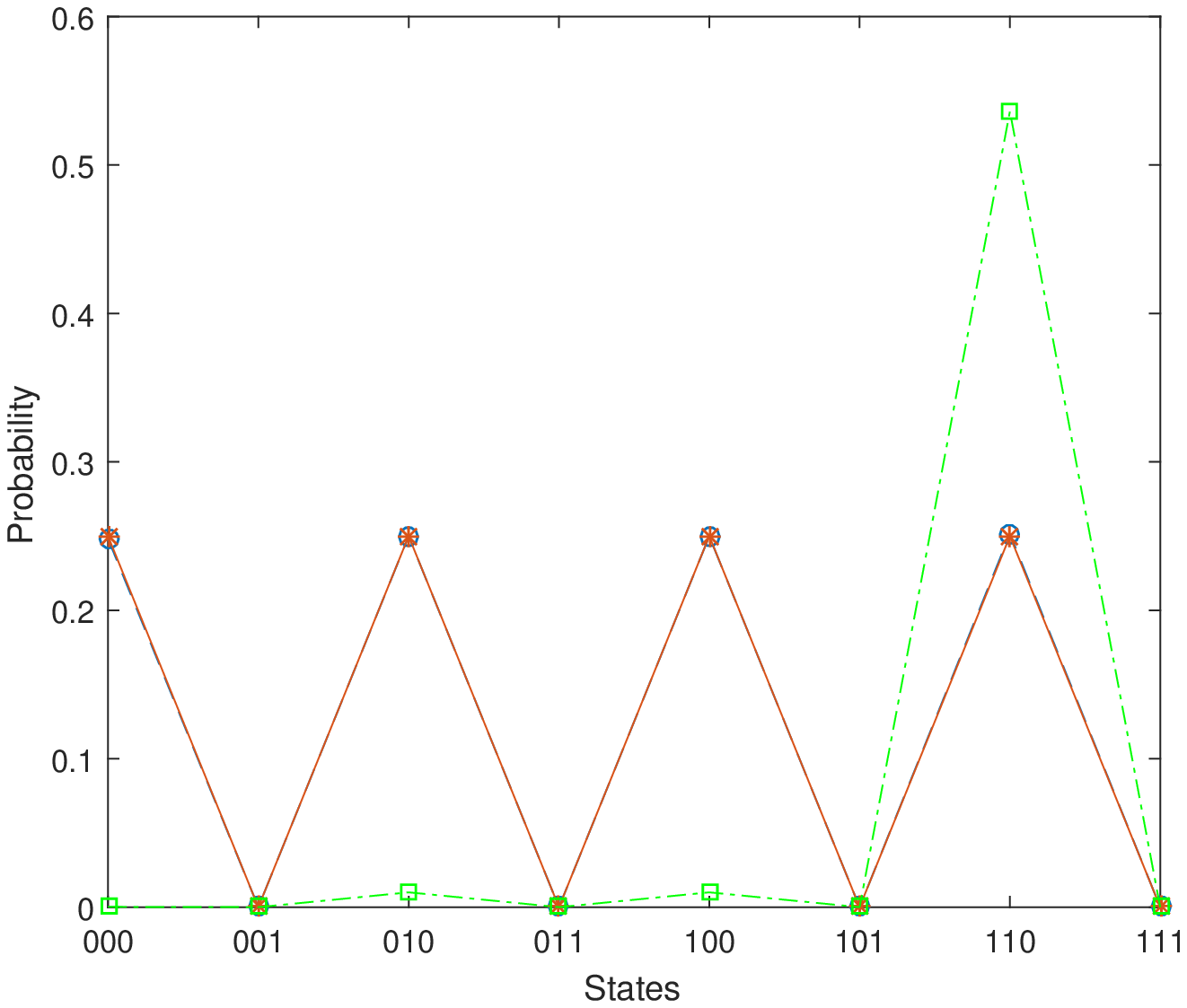}
		\caption{\emph{Even State}\\}
		\label{fig:Re}
	\end{subfigure}
	\hfill
	\begin{subfigure}[b]{0.3\textwidth}
		\centering
		\includegraphics[height=2in]{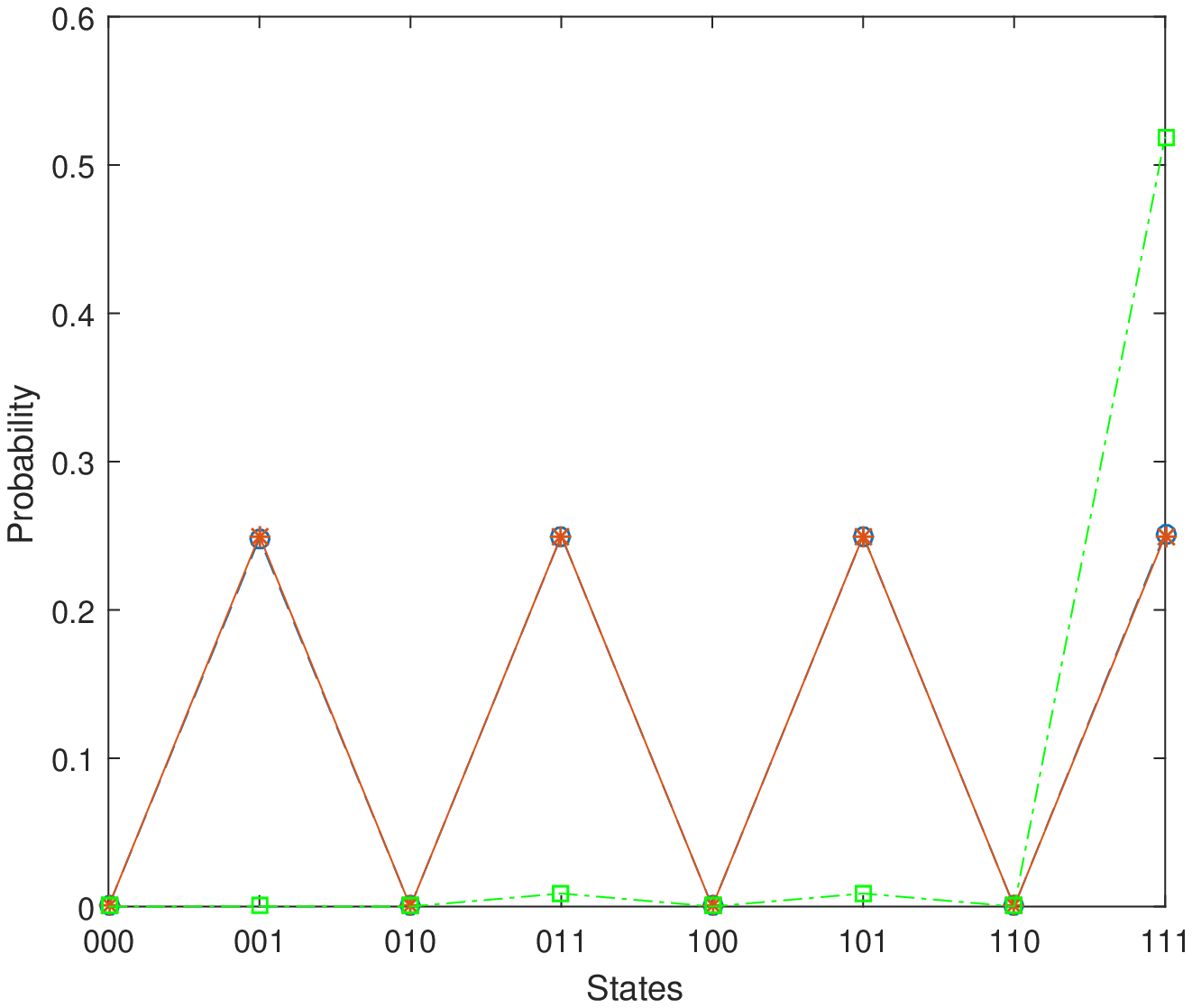}
		\caption{\emph{Odd State}\\}
		\label{fig:Rf}
	\end{subfigure}
	\hfill
	\begin{subfigure}[b]{0.3\textwidth}
		\centering
		\includegraphics[height=2in]{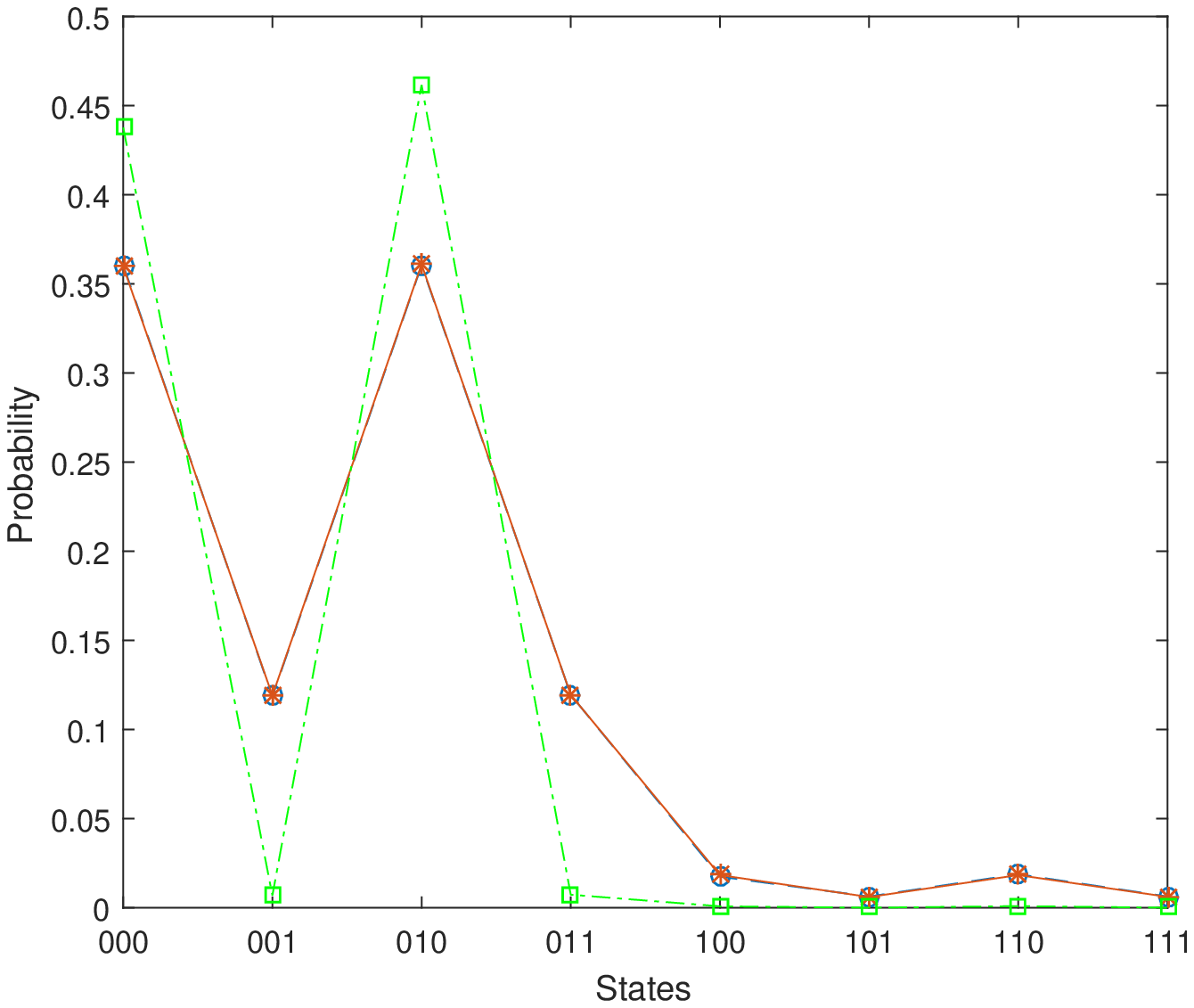}
		\caption{\emph{Random State}}
		\label{fig:Rg}
	\end{subfigure}
	\hfill
	\begin{subfigure}[b]{0.4\textwidth}
		\centering
		\includegraphics[height=2in]{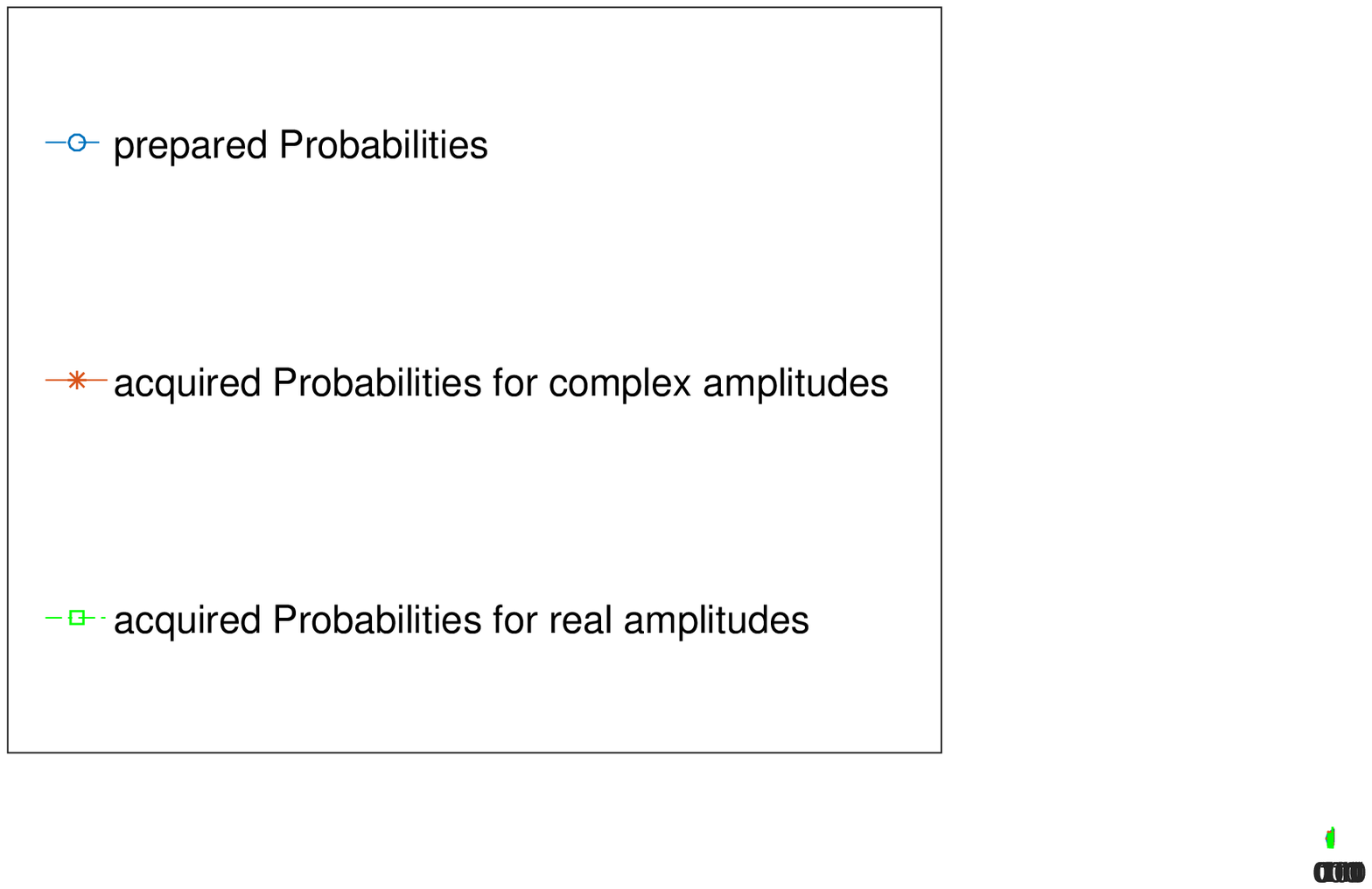}
		
	\end{subfigure}
	\caption{The different results for prepared and acquired probabilities where the blue line represents the prepared probabilities, the red line represents the acquired probabilities for complex prepared amplitudes, and the green line represents the acquired probabilities for real prepared amplitudes.}
	\label{fig:result}
\end{figure}

The plots in figure \ref{fig:result} illustrate the relation between the acquired probabilities and the prepared probabilities in each of the different test cases. The blue line in the plots represents the prepared probabilities while the red line represents the acquired probabilities for complex prepared amplitudes and the green line for real amplitudes. 
In figures \ref{fig:Ra},\ref{fig:Rc},\ref{fig:Rd},\ref{fig:Re} and\ref{fig:Rf} the blue line matches the red line with relative error between the two line in range between $10^{-4}$ and $10^{-11}$ when the values of the prepared amplitudes are complex. On the other hand, a noticeable difference between the blue and the green plots has been shown in figures \ref{fig:Rc},and \ref{fig:Rd} in the case of using prepared amplitudes with real values. However, when the prepared amplitudes complex values, the matching is less than $10^{-6}$  as shown in figures \ref{fig:Rc} and \ref{fig:Rd}. 

In summary, the experimental results showed that the proposed method is capable to accurately prepare both complete and incomplete superpositions. Complete superposition are shown in figures \ref{fig:Ra},\ref{fig:Rc},\ref{fig:Rd}and \ref{fig:Rg}, whereas, incomplete superpositions where some states have amplitudes zero as shown in figures \ref{fig:Rb}, \ref{fig:Re} and \ref{fig:Rf}.

\section{Comparison with Related Work}\label{5}
The complexity of a quantum circuit is measured by: the circuit depth (total number of used gates), number of auxiliary qubits and the number of single two qubits controlled gates in the circuit (e.g., CNOT and $C_k$). 
Many related work suggested different schemes for the preparation of quantum states \cite{12,16,25,19,27,20,21}. All circuits proposed in \cite{24,26,27} have circuit depths of exponential complexity in the number of qubits. However, the circuits presented in \cite{12,16,25,19} have circuit depths of polynomial complexity in the number of qubits. In this paper, we propose a method that uses circuits with circuit depths of linear complexity.

Regarding the complexity based on the number of auxiliary qubits, \cite{19, 25} introduced method of complexity $\mathcal{O}(n)$ of auxiliary qubits, whereas the proposed method required only one auxiliary qubit with complexity of $\mathcal{O}(1)$.

When comparing the complexity of this method based on the number of two qubits controlled gates, the work in \cite{24} presented a quantum circuit with
$2^{n+2}-4n-4$ CNOT gates where $n$ is the number of qubits. The quantum circuit presented in \cite{26}, required $2^{n+1}-2n$ CNOT gates. In \cite{27}, a quantum circuit with Universal Gate was introduced with $\frac{23}{24}2^n$ CNOT gates.
In \cite{19} the superposition for 2-qubits was generated using 3 CNOT gates and using 13 CNOT gates for 3-qubits gates. The proposed method requires only $n$ $C_k$ for n-qubits circuits.

\begin{table}[H]
	\caption{Comparison between the complexity of the proposed method and the other algorithms where $n$ is number of qubits.}
	\centering
	\begin{tabular}{c| c |c} 
		\hline 
		Algorithm & Circuit Depth & No.of auxiliary qubit    \\
		\hline
		\emph{Equal State} \cite{12}& $\mathcal{O}(n^3)$ & - \\ [1ex] 
		\hline 
		\emph{Prime State} \cite{16}& $\mathcal{O}(n^2)$ & - \\ [1ex] 
		\hline
		\emph{Universal Gate} \cite{27}  & $\mathcal{O}(2^n)$& -\\ [1ex] 
		\hline
		\emph{The Sequential Algorithm} \cite{25} &  $\mathcal{O}(n^2)$&  $\mathcal{O}(n)$\\ [1ex] 
		\hline
		\emph{QAE} \cite{19} & $\mathcal{O}(n)$&$\mathcal{O}(n)$\\ [1ex] 
		\hline
		\emph{The Proposed Method} &  $\mathcal{O}(n)$&$\mathcal{O}(1)$\\ [1ex] 
		\hline
	\end{tabular}
	\label{table2}
\end{table}

Table ~\ref{table2} summarizes the complexity of the proposed circuit compared to the complexity of other circuits introduced in literature. 
It is obvious the complexity of the proposed circuit does not exceed $\mathcal{O}(n)$ circuit depth and $\mathcal{O}(1)$ auxiliary qubit, whereas the circuit depth of the circuits proposed in the related work was at least of complexity $\mathcal{O}(n^2)$ and number of auxiliary qubits was $\mathcal{O}(n)$.

\section{Conclusion} \label{c6}
Data Encoding is usually the first step in any quantum algorithm, where the successful preparation of the required quantum superposition leads to the success of the quantum algorithms in terms of the speed-up over classical algorithms and/or the probability of success to get the correct results. Preparation of a uniform superposition is a trivial task using Walsh-Hadamard transform, where all the quantum states appear in the superposition with equal amplitude and so probability. The preparation of a non-uniform superposition where the quantum states have different amplitudes, namely an incomplete superposition where certain states should be missing, i.e. with zero probability, is a challenging problem. Many methods have been proposed, where each method is proposed to be used to prepare a certain superposition.

In this paper, an $n$-qubits variational quantum circuit has been proposed that uses $n$ partial negation operators and $n$ controlled partial negation operators to prepare an arbitrary superposition. The proposed method can be used to prepare an arbitrary quantum superposition with high accuracy in $\mathcal{O}(n)$ steps. It has been shown that the proposed method has be used successfully to prepare special quantum superpositions proposed in literature. The proposed method takes the acquired amplitudes as an input, calculate the unknown parameters of the variational quantum circuit by solving a system of nonlinear equations using Levenberg-Marquardt algorithm.  
The proposed method can be extended to prepare an arbitrary quantum superposition over n-qubits. This work can be extended by exploring the benefits of using more complex partial negation operators and to enhance the speed and the accuracy of calculating the unknow parameters of the variational quantum circuit. 

\section{Conclusion} \label{5}
In this paper, the quantum state prepares by using the partial negation operator ($r^{th}$ root) to create incomplete and complete superposition. The quantum states are used in many applications such as quantum information processing.  we generate quantum the circuit to prepare most of the distribution probability.
For the previous algorithms, the superposition preparation is created by the $H$ gate, this gate creates a perfect superposition with equal distribution probability for all states.    
A new quantum algorithm has been proposed to generate value of the $r^{th}$ root operator. Changing the $K$ gates presents different probability of the exact amplitude.
 The resulting values of error for 100 test random amplitude with 3 qubits, reduces the average of error=$6.9912e-04$. 
The previous algorithms uses the $H$ gate with complexity equal to $O(n^2)$. However, the complexity of the proposed algorithm is $O(n)$. 
 The proposed algorithm proposes state preparation,together with a circuit technique, to Develop the preparation of state preparation. For future work, the main problem is design the circuit of the set of quantum states that can be prepared in linear time complexity.
\section*{Acknowledgment}
This paper was supported financially by the Academy of Scientific Research and Technology (ASRT), Egypt,
under initiatives of Science Up Faculty of Science (Grant No 6558). (ASRT) is the 2nd affiliation of this research.\\
\bibliographystyle{ieeetr}
\bibliography{lib}

\begin{thebibliography}{10}

\bibitem{1}
E.~Rieffel and W.~Polak, ``An introduction to quantum computing for
  non-physicists,'' {\em ACM Computing Surveys (CSUR)}, vol.~32, no.~3,
  pp.~300--335, 2000.

\bibitem{2}
M.~A. Nielsen and I.~Chuang, ``Quantum computation and quantum information,''
  2002.

\bibitem{3}
A.~Ekert, P.~Hayden, and H.~Inamori, ``Basic concepts in quantum computation,''
  in {\em Coherent atomic matter waves}, pp.~661--701, Springer, 2001.

\bibitem{7}
A.~Chatterjee, ``Introduction to quantum computation,'' {\em arXiv preprint
  quant-ph/0312111}, 2003.

\bibitem{4}
J.~A. Miszczak, ``Models of quantum computation and quantum programming
  languages,'' {\em arXiv preprint arXiv:1012.6035}, 2010.

\bibitem{28}
P.~Chao-Yang, Z.~Zheng-Wei, C.~Ping-Xing, and G.~Guang-Can, ``Design of quantum
  vq iteration and quantum vq encoding algorithm taking o (n 1/2) steps for
  data compression,'' {\em Chinese Physics}, vol.~15, no.~3, p.~618, 2006.

\bibitem{29}
K.~K. Sabapathy, H.~Qi, J.~Izaac, and C.~Weedbrook, ``Production of photonic
  universal quantum gates enhanced by machine learning,'' {\em Physical Review
  A}, vol.~100, no.~1, p.~012326, 2019.

\bibitem{5}
A.~Chalumuri, R.~Kune, and B.~Manoj, ``Training an artificial neural network
  using qubits as artificial neurons: a quantum computing approach,'' {\em
  Procedia Computer Science}, vol.~171, pp.~568--575, 2020.

\bibitem{8}
L.~K. Grover, ``A fast quantum mechanical algorithm for database search,'' in
  {\em Proceedings of the twenty-eighth annual ACM symposium on Theory of
  computing}, pp.~212--219, 1996.

\bibitem{9}
G.-L. Long, ``Grover algorithm with zero theoretical failure rate,'' {\em
  Physical Review A}, vol.~64, no.~2, p.~022307, 2001.

\bibitem{17}
P.~W. Shor, ``Algorithms for quantum computation: discrete logarithms and
  factoring,'' in {\em Proceedings 35th annual symposium on foundations of
  computer science}, pp.~124--134, Ieee, 1994.

\bibitem{23}
A.~W. Harrow, A.~Hassidim, and S.~Lloyd, ``Quantum algorithm for linear systems
  of equations,'' {\em Physical review letters}, vol.~103, no.~15, p.~150502,
  2009.

\bibitem{10}
J.~Sang, S.~Wang, and Q.~Li, ``A novel quantum representation of color digital
  images,'' {\em Quantum Information Processing}, vol.~16, no.~2, p.~42, 2017.

\bibitem{24}
M.~Mottonen, J.~J. Vartiainen, V.~Bergholm, and M.~M. Salomaa, ``Transformation
  of quantum states using uniformly controlled rotations,'' {\em arXiv preprint
  quant-ph/0407010}, 2004.

\bibitem{26}
V.~V. Shende, S.~S. Bullock, and I.~L. Markov, ``Synthesis of quantum-logic
  circuits,'' {\em IEEE Transactions on Computer-Aided Design of Integrated
  Circuits and Systems}, vol.~25, no.~6, pp.~1000--1010, 2006.

\bibitem{27}
M.~Plesch and {\v{C}}.~Brukner, ``Quantum-state preparation with universal gate
  decompositions,'' {\em Physical Review A}, vol.~83, no.~3, p.~032302, 2011.

\bibitem{12}
Q.~Yu, Y.~Zhang, J.~Li, H.~Wang, X.~Peng, and J.~Du, ``Generic preparation and
  entanglement detection of equal superposition states,'' {\em SCIENCE CHINA
  Physics, Mechanics \& Astronomy}, vol.~60, no.~7, p.~070313, 2017.

\bibitem{19}
A.~C. Vazquez and S.~Woerner, ``Efficient state preparation for quantum
  amplitude estimation,'' {\em Physical Review Applied}, vol.~15, no.~3,
  p.~034027, 2021.

\bibitem{25}
X.-M. Zhang, M.-H. Yung, and X.~Yuan, ``Low-depth quantum state preparation,''
  {\em arXiv preprint arXiv:2102.07533}, 2021.

\bibitem{16}
J.~I. Latorre and G.~Sierra, ``Quantum computation of prime number functions,''
  {\em arXiv preprint arXiv:1302.6245}, 2013.

\bibitem{22}
A.~D. Vos and S.~D. Baerdemacker, ``Symmetry groups for the decomposition of
  reversible computers, quantum computers, and computers in between,'' {\em
  Symmetry}, vol.~3, no.~2, pp.~305--324, 2011.

\bibitem{13}
A.~Younes, ``Reading a single qubit system using weak measurement with variable
  strength,'' {\em Annals of Physics}, vol.~380, pp.~93--105, 2017.

\bibitem{15}
J.~J. Mor{\'e}, ``The levenberg-marquardt algorithm: implementation and
  theory,'' in {\em Numerical analysis}, pp.~105--116, Springer, 1978.

\bibitem{20}
S.~Woerner and D.~J. Egger, ``Quantum risk analysis,'' {\em npj Quantum
  Information}, vol.~5, no.~1, pp.~1--8, 2019.

\bibitem{21}
D.~S. Abrams and C.~P. Williams, ``Fast quantum algorithms for numerical
  integrals and stochastic processes,'' {\em arXiv preprint quant-ph/9908083},
  1999.

\end{thebibliography}


\end{document}